\documentclass[useAMS,usenatbib]{mn2e}
\usepackage{epsfig}
\usepackage{multirow}
\usepackage{enumerate}

\title[Investigation of the emission radii of...]
{Investigation of the emission radii of kHz QPOs for the accreting millisecond X-Ray pulsars, Atoll and Z sources}

\author[D. H. Wang et al.]
{D. H. Wang$^{1,2,3}\thanks{huazai05105220@163.com}$, L. Chen$^2$, C. M. Zhang$^3$, Y. J. Lei$^3$, J. L. Qu$^4$, L. M. Song$^4$
\\
$^1$School of Physics and Electronic Science, Guizhou Normal University, Guiyang, 550001, China\\
$^2$Astronomy Department, Beijing Normal University, Beijing, 100875, China\\
$^3$National Astronomical Observatories, Chinese Academy of Sciences, Beijing, 100012, China\\
$^4$Institute of High Energy Physics, Chinese Academy of Sciences, Beijing, 100049, China\\
}
\date{Released 2002 Xxxxx XX}

\pagerange{\pageref{firstpage}--\pageref{lastpage}} \pubyear{2002}

\begin{document}

\maketitle

\label{firstpage}

\begin{abstract}

We infer the emission positions of twin kilohertz quasi-periodic oscillations (kHz QPOs) in neutron star low mass X-ray binaries (NS-LMXBs) based on the Alfv\'en wave oscillation model (AWOM). For most sources, the emission radii of kHz QPOs cluster around a region of $16-19$\,km with the assumed NS radii of 15\,km. Cir X-1 has the larger emission radii of $23\sim38$\,km than those of the other sources, which may be ascribed to its large magnetosphere-disk radius or strong NS surface magnetic field. SAX J1808.4-3658 is also a particular  source with the relative large emission radii of kHz QPOs of $20\sim23$\,km, which may be due to its large inferred NS radius of $18\sim19$\,km. The emission radii of kHz QPOs for all the sources are larger than the NS radii, and the possible explanations of which are presented. The similarity of the emission radii of kHz QPOs ($\sim16-19$\,km) for both the  low/high  luminosity Atoll/Z sources is found, which indicates that both sources share the similar magnetosphere-disk radii.

\end{abstract}


\section{Introduction}

The launch of the Rossi X-ray Timing Explorer (RXTE) has led to the discovery of kilohertz quasi-periodic oscillations (kHz QPOs) in neutron star (NS) low mass X-ray binaries (LMXBs) \citep{van der Klis96,Strohmayer96}. These high-frequency QPOs usually appear in pairs (upper $\nu_2$ and lower $\nu_1$) at frequencies from a few hundred Hz to more than 1\,kHz, as shown in Atoll sources, Z sources (see \citealt{Hasinger89} for the definition of Atoll and Z sources) and the accreting millisecond X-Ray pulsars (AMXPs) (see \citealt{van der Klis06} for a review). It is suggested that kHz QPOs might be used to test the General Relativity in a strong gravitational field regime \citep{Miller98,Stella99a}, and help to constrain the NS $Mass-Radius$ relation \citep{Miller98}.

The frequencies of the kHz QPOs are strongly correlated with other timing and spectral features, such as the photon indexes of the power-law component of the energy spectrum \citep{Kaaret98}, the noise features \citep{Ford98c},
the positions in the X-ray color-color diagram (e.g., \citealt{Wijnands97b}), the X-ray luminosities \citep{Mendez99b,Ford00}, the low frequency QPOs \citep{Psaltis99,Belloni02}, where the high-/low-frequency correlation is similar to those in black hole candidates and white dwarf cataclysmic variables (e.g. \citealt{Psaltis99,Belloni02,Warner02,Mauche02}). Besides, the quality factors and $rms$ amplitudes of the kHz QPOs can vary as a function of the QPO frequencies (e.g., \citealt{Mendez01,Barret05b}).

There is currently no consensus as to the origin of these QPOs. Some interpretations, such as
the beat frequency model by \citet{Miller98,Lamb01} and the resonance model by \citet{Kluzniak01,Abramowicz03a,Abramowicz03b}, encounter difficulties when interpreting the following observational characteristics of the kHz QPOs \citep{Belloni05,Belloni07}.
While the relativistic precession model \citep{Stella99a,Stella99b} fits the $\Delta\nu~vs.\nu_2~$ relation to the kHz QPO data quite well. However, it requires the NS mass of 2 solar mass ($\rm M_\odot$) \citep{van der Klis06}, which is relative higher than the actual measurement results \citep{Wang13a}.

The Alfv\'en wave oscillation model (AWOM) by \citet{Zhang04} makes a good description of the kHz QPO data \citep{Belloni07}. The model predicts the relative emission positions of the kHz QPOs ($X\equiv R/r$, the ratio between the NS radius and the kHz QPO emission radius), which can help to infer the emission radii of kHz QPOs, then further infer the magnetosphere-accretion disk structure and evaluate the possibility of the kHz QPO mechanism.
\begin{table*}
\begin{minipage}{122mm}
\caption{The frequency ranges and the emission radii of kHz QPOs.}
\begin{tabular}{@{}lccccl@{}}
\hline
\noalign{\smallskip}
Source (25) & $\nu_1$ & $\nu_2$ & $X$ & $r^\dag$ &  References \\
 & (Hz) & (Hz) & ($\equiv R/r$) & (km) & \\
\noalign{\smallskip}
\hline
\noalign{\smallskip}
AMXP (2) \\
SAX J1808.4-3658 & $499\sim504$ & $685\sim694$ & $0.87\sim0.88$ & -- & 1 \\
XTE J1807.4-294 & $106\sim370$ & $337\sim587$ & $0.49\sim0.83$ & $18.18\sim30.51$ & 2 \\
\noalign{\smallskip}
\hline
\noalign{\smallskip}
Atoll (15) \\
4U 0614+09 & $153\sim843$ & $449\sim1162$ & $0.52\sim0.88$ & $17.02\sim28.72$ & 3 \\
4U 1608-52 & $473\sim867$ & $799\sim1104$ & $0.77\sim0.92$ & $16.32\sim19.49$ & 4 \\
4U 1636-53 & $529\sim979$ & $823\sim1228$ & $0.81\sim0.93$ & $16.15\sim18.50$ & 5 \\
4U 1702-43 & 722 & $1055$ & $0.84$ & $17.78$ & 6\\
4U 1705-44 & $776$ & $1074$ & $0.87$ & $17.21$ & 7 \\
4U 1728-34 & $308\sim894$ & $582\sim1183$ & $0.71\sim0.90$ & $16.69\sim21.02$ & 8 \\
4U 1735-44 & $641\sim900$ & $982\sim1149$ & $0.82\sim0.91$ & $16.44\sim18.32$ & 9 \\
4U 1820-30 & $764\sim796$ & $1055\sim1072$ & $0.86\sim0.89$ & -- & 10 \\
4U 1915-05 & $224\sim707$ & $514\sim1055$ & $0.62\sim0.83$ & $18.02\sim24.09$ & 11 \\
Aql X-1 & $795\sim803$ & $1074\sim1083$ & $0.88$ & $16.95\sim16.96$ & 12 \\
IGR J17191-2821 & $681\sim870$ & $1037\sim1185$ & $0.82\sim0.88$ & $17.05\sim18.31$ & 13 \\
KS 1731-260 & $898$ & $1159\sim1183$ & $0.90\sim0.91$ & $16.53\sim16.71$ & 14 \\
SAX J1750.8-2900 & $936$ & $1253$ & $0.89$ & $16.88$ & 15 \\
XTE J1701-407 & $745$ & $1150$ & $0.82$ & $18.40$ & 16 \\
XTE J2123-058 & $847\sim871$ & $1102\sim1141$ & $0.89\sim0.90$ & $16.61\sim16.77$ & 17 \\
\noalign{\smallskip}
\hline
\noalign{\smallskip}
Z (8) \\
Cir X-1 & $56\sim226$ & $229\sim505$ & $0.40\sim0.64$ & $23.36\sim37.58$ & 18 \\
Cyg X-2 & $516$ & $862$ & $0.77$ & -- & 19 \\
GX 5-1 & $156\sim662$ & $478\sim888$ & $0.51\sim0.89$ & $16.90\sim29.69$ & 20 \\
GX 17+2 & $475\sim830$ & $759\sim1079$ & $0.79\sim0.91$ & $16.47\sim19.05$ & 21 \\
GX 340+0 & $197\sim565$ & $535\sim840$ & $0.55\sim0.83$ & $17.97\sim27.18$ & 22 \\
GX 349+2 & $715$ & $985$ & $0.87$ & $17.16$ & 23 \\
Sco X-1 & $532\sim902$ & $842\sim1143$ & $0.80\sim0.92$ & $16.34\sim18.74$ & 24 \\
XTE J1701-462 & $502\sim651$ & $761\sim945$ & $0.82\sim0.87$ & $17.33\sim18.34$ & 25 \\
\noalign{\smallskip}
\hline
\noalign{\smallskip}
\end{tabular}
\label{data_All}
\end{minipage}
\begin{tabular}{@{}l@{}}
\begin{minipage}{122mm}
Notes: The second and third columns show the frequency ranges of the lower- and upper-kHz QPOs. The forth column shows the range of the inferred position parameter $X(\equiv R/r$, see equation (\ref{X})) based on AWOM. The fifth column shows the range of the emission radii of kHz QPOs.\\
$^\dag$: The NS radii are assumed as 15\,km.\\
References:
1. \citealt{van Straaten05},  \citealt{Wijnands03};
2. \citealt{Linares05}, \citealt{Zhang06b};
3. \citealt{van Straaten00}, \citealt{van Straaten02}, \citealt{Boutelier09};
4. \citealt{van Straaten03}, \citealt{Barret05}, \citealt{Jonker00a}, \citealt{Mendez98};
5. \citealt{Altamirano08}, \citealt{Wijnands97a}, \citealt{Bhattacharyya10}, \citealt{Di Salvo03}, \citealt{Jonker00a}, \citealt{Jonker02a}, \citealt{Lin11};
6. \citealt{Markwardt99};
7. \citealt{Ford98a};
8. \citealt{Di Salvo01}, \citealt{van Straaten02}, \citealt{Strohmayer96}, \citealt{Migliari03}, \citealt{Jonker00a}, \citealt{Mendez99};
9. \citealt{Wijnands98c}, \citealt{Ford98b};
10. \citealt{Smale97};
11. \citealt{Boirin00};
12. \citealt{Barret08};
13. \citealt{Altamirano10};
14. \citealt{Wijnands97};
15. \citealt{Kaaret02};
16. \citealt{Strohmayer08};
17. \citealt{Tomsick99}, \citealt{Homan99};
18. \citealt{Boutloukos06};
19. \citealt{Wijnands98a};
20. \citealt{Wijnands98b}, \citealt{Jonker02b};
21. \citealt{Homan02}, \citealt{Wijnands97b};
22. \citealt{Jonker02b}, \citealt{Wijnands98a}, \citealt{Jonker98};
23. \citealt{O'Neill02};
24. \citealt{van der Klis97}, \citealt{van der Klis96}, \citealt{Lin11}, \citealt{Mendez00};
25. \citealt{Homan07}, \citealt{Homan10}, \citealt{Sanna10}.
\end{minipage}
\end{tabular}
\end{table*}
\begin{table*}
\begin{minipage}{160mm}
\caption{The emission radii of kHz QPOs for the sources with measured NS masses$^\ddag$.}
\begin{tabular}{@{}lccccccc@{}}
\hline
\noalign{\smallskip}
Source (3) & Measured NS mass & Ref & $A$ & $R$ & $r$ & $R_{\rm ISCO}$ \\
 & ($\rm M_\odot$) & & ($\propto{\rho^{1/2}}$) & (km) & (km) & (km) \\
\noalign{\smallskip}
\hline
\noalign{\smallskip}
Cyg X-2 & $1.71\pm0.21$ & 1 & $0.62\pm0.04$ & $16.4\pm1.02$ & $21.23\pm1.60$ & $15.12\pm1.86$ \\
SAX J1808.4-3658 & $<1.4$ & 2 & $0.43\pm0.01$ & $18\sim20^\|$ & $20\sim23^\|$ & $9\sim12^\|$ \\
XTE 1820-30 & $1.29^{+0.19}_{-0.07}$ & 3 & $0.65\pm0.01$ & $14.5^{+0.7}_{-0.3}$ & $16.2^{+0.8}_{-0.4}\sim16.7^{+0.9}_{-0.4}$ & $11.4^{+1.7}_{-0.6}$ \\
\noalign{\smallskip}
\hline
\noalign{\smallskip}
\end{tabular}
\label{data_mass}
\end{minipage}
\\
\begin{tabular}{@{}l@{}}
\begin{minipage}{160mm}
Notes: The second column shows the measured NS masses. The forth column shows the NS density parameter $A(\equiv\sqrt{\frac{M}{M_\odot}\frac{1}{R^3_6}}$, see equation (\ref{A})) inferred by the twin kHz QPO frequencies in Table \ref{data_All}. The fifth column shows the NS radii inferred by the measured NS masses and $A$. The sixth column shows the range of the emission radii of kHz QPOs inferred by the NS radii and position parameter $X$ in Table \ref{data_All}. The last column shows the ISCO radii (i.e. $R_{\rm ISCO}=3R_{\rm s}=6GM/{\rm c}^2)$.\\
$^\ddag$: See Table \ref{data_All} for the twin kHz QPOs and position parameter $X$.\\
$^\|$: The lower limit of the NS mass is adopted as $1\,\rm M_\odot$(see \citealt{Miller02,Zhang11}). \\
Reference:
1. \citealt{Casares10};
2. \citealt{Elebert09a};
3. \citealt{Shaposhnikov04}.
\end{minipage}
\end{tabular}
\end{table*}

In this paper we analyze the emission positions of kHz QPOs based on AWOM: In $\S$ 2, we introduce the AWOM and show the kHz QPO data used in the paper. In $\S$ 3 we infer the emission radii of kHz QPOs and analyze the results for the particular sources. In $\S$ 4 we make discussions and conclusions.

\section{The model and data}

\subsection{The magnetosphere-disk radius}

It is thought that kHz QPOs reflect the motion of matter in orbit at some preferred radius in the accretion
disk around the neutron star in LMXBs \citep{Miller98,Stella99a,Osherovich99,Lamb01,Zhang04}.
And in order to consider the interaction between the accretion flow and the magnetic field of NS, we introduce the magnetosphere-disk radius $r_{\rm m}$, where the kinetic energy of the free-falling gas becomes comparable to the magnetic energy of the NS magnetosphere \citep{Shapiro83}:

\begin{equation}
r_{\rm m}=\xi r_{\rm A}=\xi(\frac{\mu^4}{2GM\dot{M}^2})^{1/7}
\label{r_m}
\end{equation}
where $\xi$ is a constant factor that is usually taken as 0.5 \citep{Ghosh79}, $r_{\rm A}$ is the Alfv\'en radius, $\mu$ is the magnetic moment of NS, $G$ is the gravitational constant, $M$ is the NS mass and $\dot{M}$ is the accretion rate at the inner disk boundary.

We assume the NS magnetic field to be dipolar:
\begin{equation}
B(r)=B_{\rm s}(\frac{R}{r})^3
\label{NS_B}
\end{equation}
where $B_{\rm s}$ is the NS surface dipole magnetic field strength, $R$ is the NS radius and $r$ is the radial distance refer to the center of the NS, and it can be seen that the magnetic field strength decreases with $r^3$. Then we adopt the accreting NS mass $M$ of 1.6\,$\rm M_\odot$ based on its standard value of 1.4\,$\rm M_\odot$ (the NS mass in LMXBs is averagely increased by about 0.2\,$\rm M_\odot$ on account of the accretion, see \citealt{Zhang11}) and assume the NS radius $R$ of 15\,km, then the magnetosphere-disk radius can be written as:
\begin{equation}
r_{\rm m}\approx7({\rm km})(\frac{B_{\rm s}}{10^8\,{\rm G}})^{4/7}(\frac{\dot{M}}{10^{18}\,{\rm g\,s^{-1}}})^{-2/7}.
\label{r_m_NS}
\end{equation}

\subsection{AWOM}
\begin{figure}
\centering
\includegraphics[width=7.5cm]{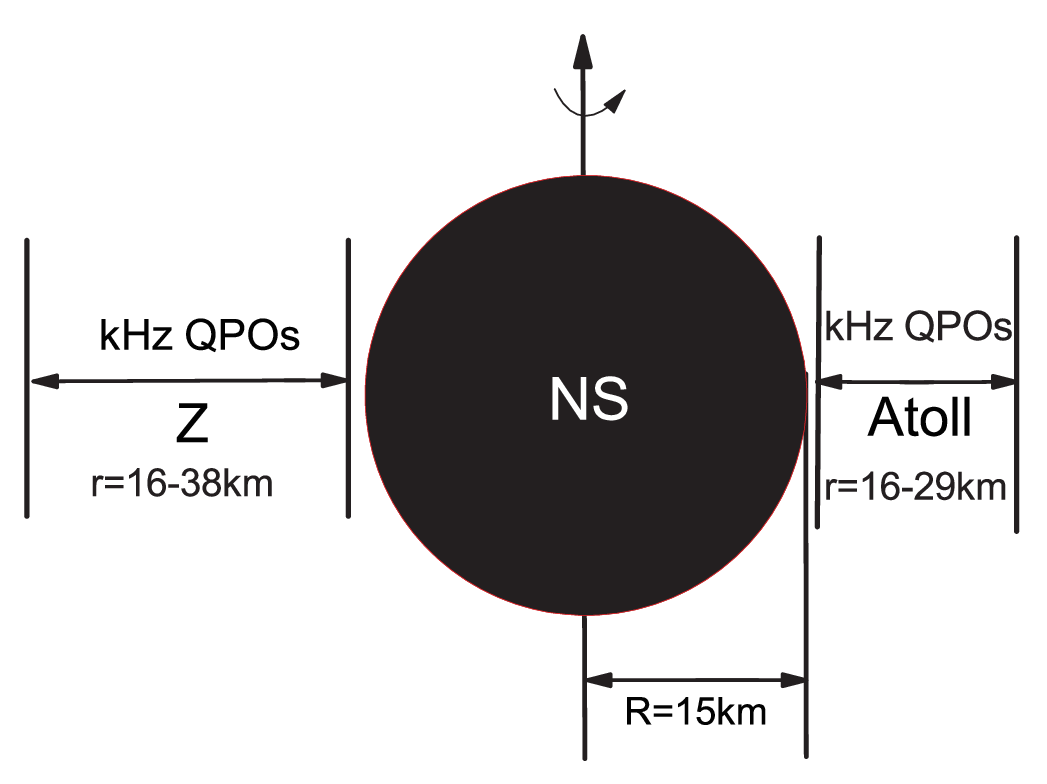}
\caption{The schematic diagram of the emission radii of kHz QPOs for Atoll and Z sources. $R$ is the NS radius that is assumed to be 15\,km, and $r$ is the radial emission radius of kHz QPOs with the range of $16-29$\,km and $16-38$\,km for Atoll and Z sources, respectively (see Table \ref{data_All} and Table \ref{data_mass} for the details).}
\label{NS_Accretion}
\end{figure}
\begin{figure}
\centering
\includegraphics[width=7.5cm]{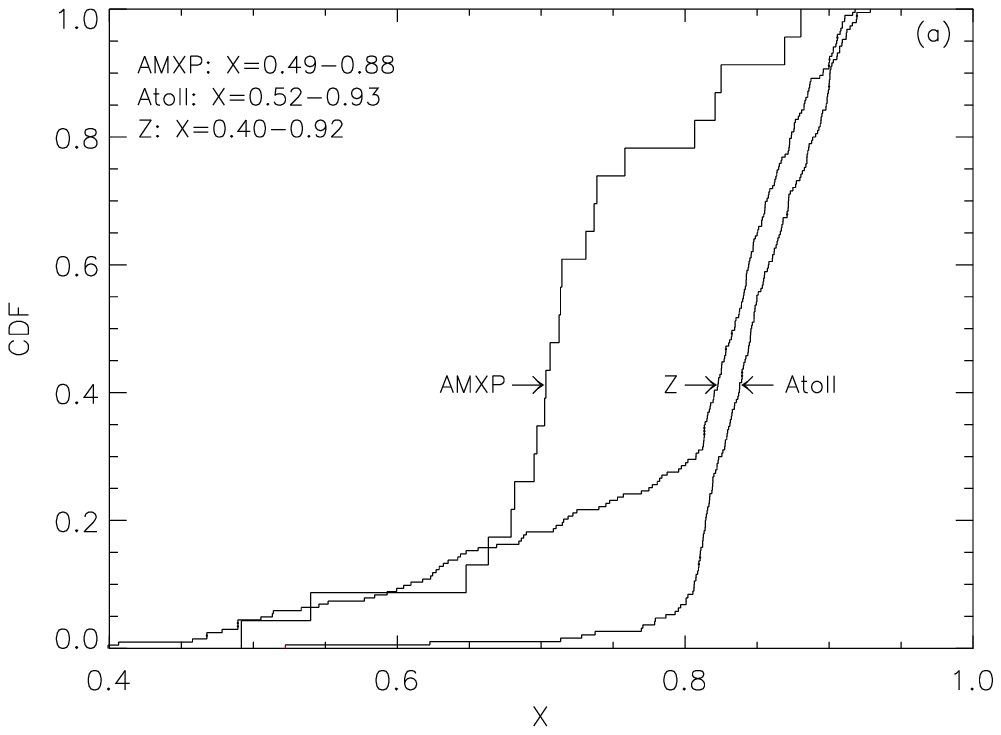}
\includegraphics[width=7.5cm]{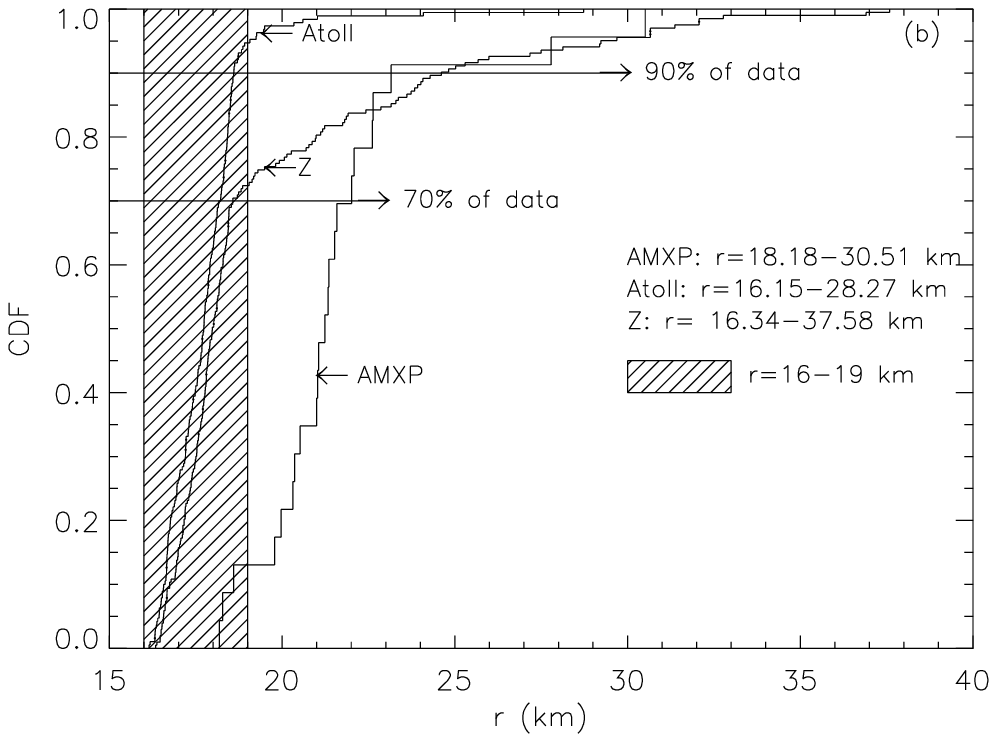}
\caption{(a) The CDF curves of the position parameter $X$ values. The abscissa and ordinate represent the $X$ values and the CDF values respectively, and the arrows indicate the curves for AMXPs ($X=0.49-0.88$), Atoll ($X=0.52-0.93$) and Z ($X=0.40-0.92$) sources. (b) Same as (a), but for the emission radius $r$ of kHz QPOs (the corresponding ranges of $r$ are $18.18-30.51$\,km, $16.15-28.27$\,km and $16.34-37.58$\,km for AMXPs, Atoll and Z sources respectively), and the shaded area shows the radius range of 16-19\,km, which contains most of the data of Atoll sources ($>90\%$) and Z sources ($>70\%$) (see Table \ref{data_All} and Table \ref{data_mass} for the details).
}
\label{r_Atoll_Z_MSP}
\end{figure}
In AWOM, the upper kHz QPO frequency $\nu_2$ is assumed as
the Keplerian orbital frequency $\nu_{\rm K}$ of the accretion flow at the preferred radius
$r$ \citep{Zhang04}:
\begin{equation}
\nu_2=\nu_{\rm K}=\sqrt{\frac{GM}{4\pi^2r^3}}
\label{nu_2}
\end{equation}
while the lower kHz QPO frequency $\nu_1$ is interpreted as the MHD Alfv\'en wave
oscillation frequency there. The model predicts the following relations between the NS mass density, the emission position of the kHz QPOs and the twin kHz QPO frequencies:
\begin{equation}
\nu_1=629({\rm Hz})A^{-2/3}\nu_{2\rm k}^{5/3}\sqrt{1-\sqrt{1-(\frac{\nu_{2\rm k}}{1.85A})^{2/3}}}
\label{nu_2_nu_1_1}
\end{equation}
\begin{equation}
A=\sqrt{\frac{M}{M_\odot}\frac{1}{R^3_6}}
\label{A}
\end{equation}
\begin{equation}
\rho=4.75\times10^{14}A^2({\rm g\,cm^{-3}})
\label{A_Fun3}
\end{equation}
\begin{equation}
\nu_2=\nu_1X^{-5/4}\sqrt{1+\sqrt{1-X}}
\label{nu_2_nu_1_2}
\end{equation}
\begin{equation}
X\equiv R/r
\label{X}
\end{equation}
where $\nu_{2\rm k}=\nu_2/1000\rm \,Hz$, $A$ is the parameter that is related to the NS mass density $\rho$, $M$ is the NS mass, $R_6=R/10^6\rm \,cm$ is the NS radius $R$ in unit of 10\,km, $X$ is the position parameter and $r$ is the emission radius of the kHz QPOs refer to the center of the NS.
According to AWOM, the NS mass density parameter $A$ and the position
parameter $X$ can be inferred by the twin kHz QPO frequencies.

\begin{figure*}
\centering
\includegraphics[width=7.5cm]{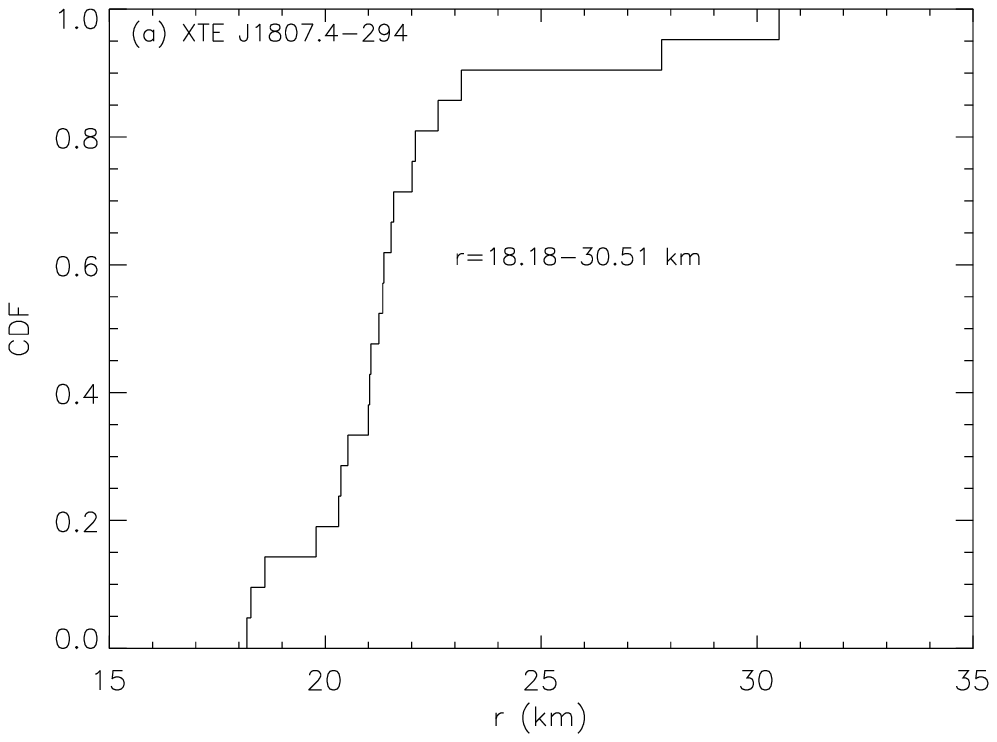}
\includegraphics[width=7.5cm]{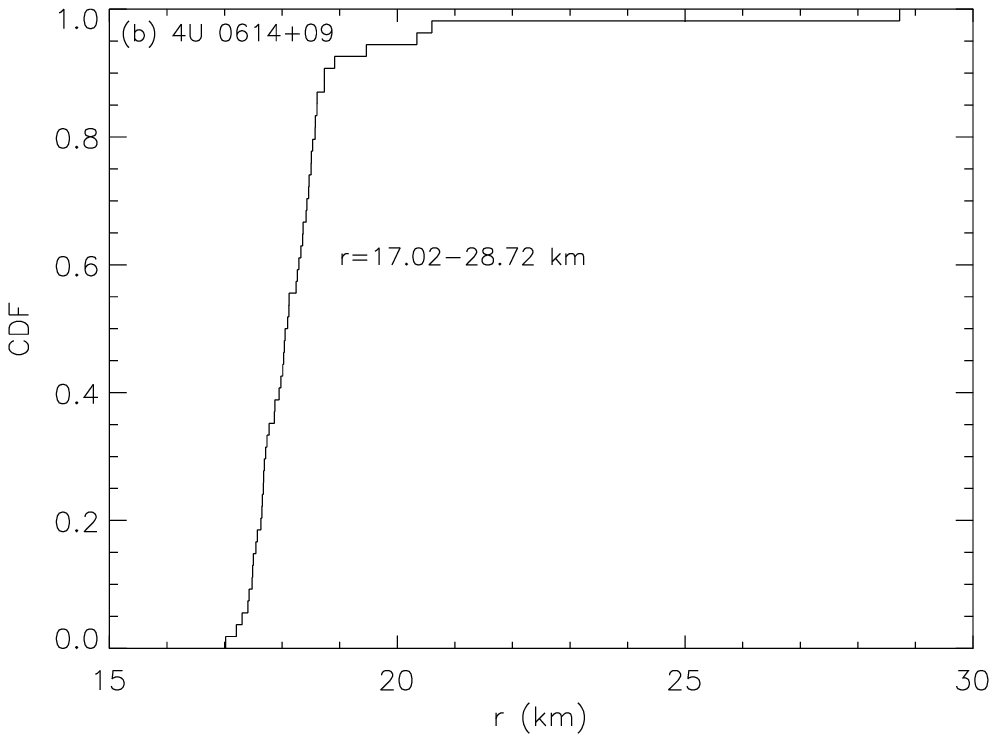}\\
\includegraphics[width=7.5cm]{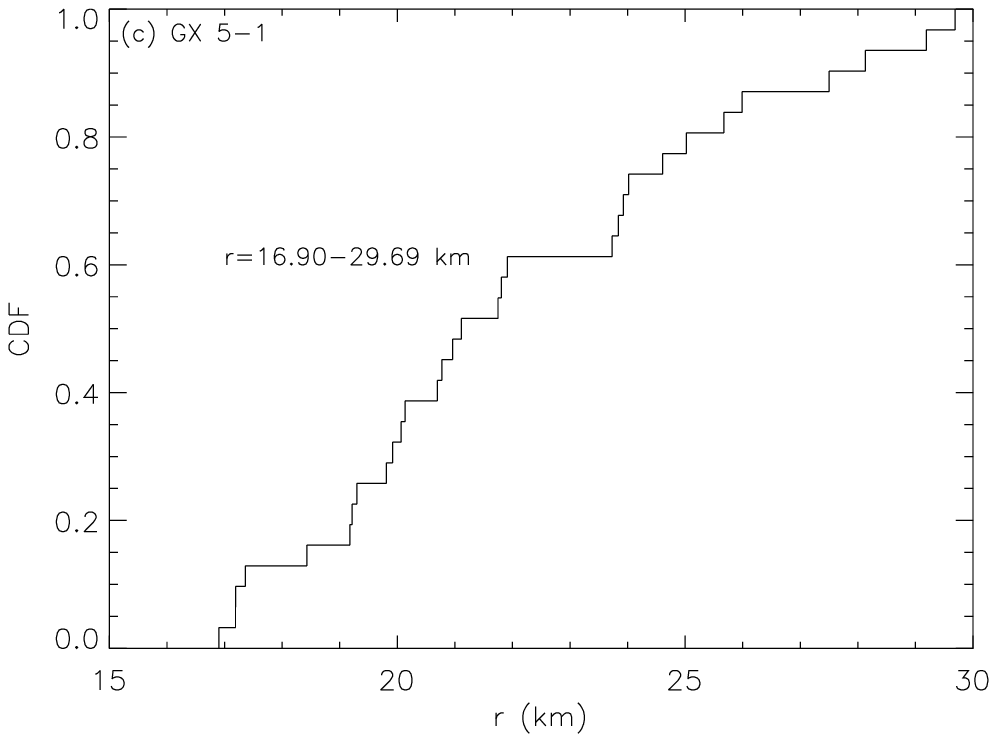}
\includegraphics[width=7.5cm]{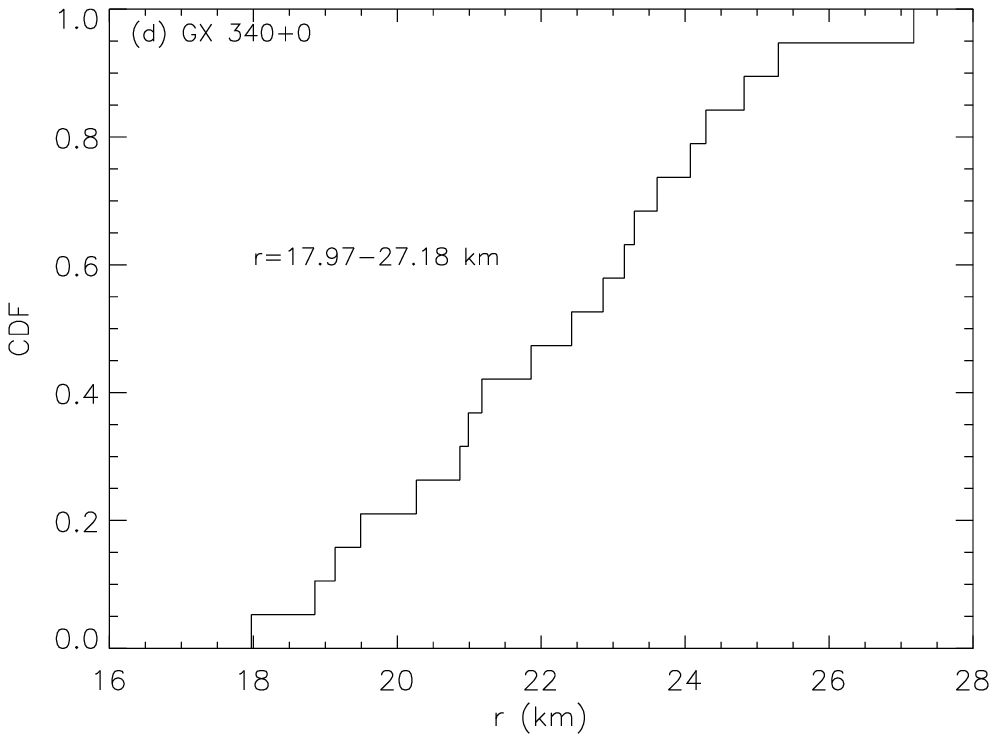}\\
\includegraphics[width=7.5cm]{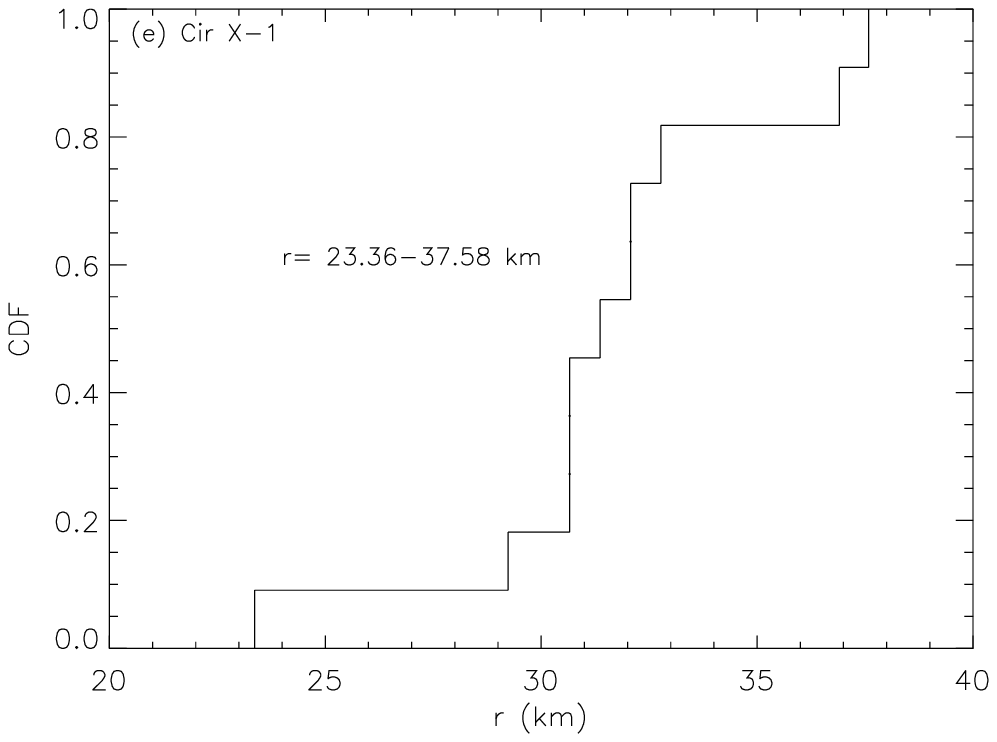}
\caption{The CDF curves of the emission radii of kHz QPOs for (a) XTE J1807.4-294 ($r=18.18-30.51$\,km), (b) 4U 0614+09 ($r=17.02-28.72$\,km), (c) GX 5-1 ($r=16.90-29.69$\,km), (d) GX 340+0 ($r=17.97-27.18$\,km), (e) Cir X-1 ($r=23.36-37.58$\,km), see Table \ref{data_All} for the details.}
\label{r_Z}
\end{figure*}

\subsection{The sample of published kHz QPO frequencies}

We search the literature for all published instances of kHz QPO frequencies, and constrain the sample to the detection of two simultaneous kHz QPO peaks.
The sample includes 416 pairs of twin kHz QPOs from 25 NS-LMXB sources, where the 23 pairs are taken from the accreting millisecond X-Ray pulsars, and the 190/203 ones from Atoll/Z sources. The adopted sources with the ranges
of the lower and upper kHz QPO frequencies and references are reported in Table \ref{data_All}. In addition, three sources in
the sample have been measured with the NS masses, which are shown in Table \ref{data_mass}.

\section{The emission positions of kHz QPOs}

We infer the emission radii of kHz QPOs based on AWOM.
For the 22 sources without the measured NS masses (see Table \ref{data_All}), we take the following calculation steps:
\begin{enumerate}[(1)]
\item For a certain pair of kHz QPO frequencies ($\nu_1$ and $\nu_2$), we first calculate the value of the position parameter $X$ by solving equation (\ref{nu_2_nu_1_2}):\\ $\nu_2=\nu_1X^{-5/4}\sqrt{1+\sqrt{1-X}}$.
\item Then we infer the corresponding emission radius $r$ of the kHz QPOs with the $X$ value by equation (\ref{X}): $r=R/X$, where the NS radius $R$ is assumed to be 15\,km.
\end{enumerate}
While for the three sources with the measured NS masses (Cyg X-2, SAX J1808.4-3658 and XTE 1820-30, see Table \ref{data_mass}), we first take the following steps to infer the NS radius $R$ based on AWOM:
\begin{enumerate}[(1)]
\item For each source, we fit the equation (\ref{nu_2_nu_1_1}):\\ $\nu_1=629({\rm Hz})A^{-2/3}\nu_{2\rm k}^{5/3}\sqrt{1-\sqrt{1-(\frac{\nu_{2\rm k}}{1.85A})^{2/3}}}$ \\to the twin kHz QPO frequencies and obtain the value of the NS density parameter $A$.
\item Then we infer the NS radius $R$ of this source with the $A$ value and the measured NS mass $M$ by solving equation (\ref{A}): $A=\sqrt{\frac{M}{M_\odot}\frac{1}{R^3_6}}$.
\end{enumerate}
Next we take the following steps to infer the emission radii of kHz QPOs for the three sources:
\begin{enumerate}[(1)]
\item For a certain pair of kHz QPO frequencies, we calculate the $X$ value with $\nu_1$ and $\nu_2$ by solving equation (\ref{nu_2_nu_1_2}): $\nu_2=\nu_1X^{-5/4}\sqrt{1+\sqrt{1-X}}$.
\item Then we infer the corresponding emission radius $r$ of the kHz QPOs with the $X$ value and the inferred NS radius $R$ by equation (\ref{X}): $r=R/X$.
\end{enumerate}

The ranges of the inferred $X$ and $r$ values for the sources without and with the measured NS masses are shown in Table \ref{data_All} and Table \ref{data_mass}, respectively. There are some aspects should be noticed:
\begin{enumerate}[(1)]
\item The emission radii of kHz QPOs of AMXPs, Atoll and Z sources are distributed in the range of $18-31$\,km, $16-29$\,km and $16-38$\,km, respectively. Fig.\ref{NS_Accretion} shows the schematic diagram of the radial ranges of the emission positions of kHz QPOs for Atoll and Z sources. It is evidence that Z sources have the larger range of the emission radii of kHz QPOs than Atoll sources, which may result from source Cir X-1 with the large emission radii of $23.36-37.58$\,km (see also Fig.\ref{r_Z}).
\item We show the cumulative distribution function (CDF) curves of the position parameter $X$ and the emission radius $r$ of kHz QPOs for AMXPs ($X=0.49-0.88$, $r=18.18-30.51$\,km), Atoll ($X=0.52-0.93$, $r=16.15-28.72$\,km) and Z ($X=0.40-0.92$, $r=16.34-37.58$\,km) sources in Fig.\ref{r_Atoll_Z_MSP}. It can be seen that most of the emission radii of kHz QPOs of Atoll sources ($>90\%$ of the data) and Z sources ($>70\%$ of the data) cluster around the region of $16-19$\,km.
\item XTE J1807.4-294 ($r=18.18-30.51$\,km), 4U 0614+09 ($r=17.02-28.72$\,km), GX 5-1 ($r=16.90-29.69$\,km) and GX 340+0 ($r=17.97-27.18$\,km) show the larger range of the emission radii of kHz QPOs from $\sim$ 17\,km to $\sim$ 30\,km (see Table \ref{data_All} for the details) than the other sources, and their corresponding CDF curves of the emission radii are shown in Fig.\ref{r_Z}.
\item We also analyze the innermost emission positions of kHz QPOs for all the sources in the sample. Fig.\ref{X_r} shows the CDF curve of the minimal emission radii ($16.15-23.36$\,km) of all the 25 sources, from which it can be seen that 22 sources share the minimal emission radii around the range of $16-19$\,km, while the other three sources, i.e. Cir X-1, Cyg X-2 and SAX J1808.4-3658, show the larger ones of about 23, 21 and 20\,km, respectively.
\end{enumerate}

\section{Discussions and Conclusions}

\begin{figure}
\centering
\includegraphics[width=7.5cm]{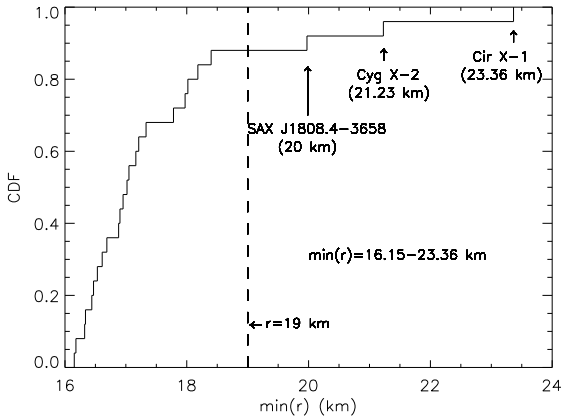}
\caption{The CDF curve of the minimal emission radii of kHz QPOs ($16.15-23.36$\,km) of all the 25 sources in Table \ref{data_All} and Table \ref{data_mass}. The dashed line shows the position of the emission radius of 19\,km and the arrows indicate the innermost emission positions of source Cir X-1 ($r_{\rm min}=23.36$\,km), Cyg X-2 ($r_{\rm min}=21.23$\,km) and SAX J1808.4-3658 ($r_{\rm min}=20$\,km) (see Table \ref{data_All} and Table \ref{data_mass} for the details).
}
\label{X_r}
\end{figure}

We investigate the emission radii of kHz QPOs for the accreting millisecond X-ray pulsars, Atoll and Z sources based on AWOM, and find that most of the emission positions of kHz QPOs cluster around the region of several kilometers away from the NS surface. The details of the conclusions are discussed and summarized as follows:
\begin{enumerate}[(1)]
\item The inferred emission radii of kHz QPOs of AMXPs, Atoll and Z sources based on AWOM are distributed in the range of $18-31$\,km, $16-29$\,km and $16-38$\,km respectively 
     (see Table \ref{data_All}, Table \ref{data_mass} and Fig.\ref{NS_Accretion} for the details). Cir X-1 has the obviously larger emission radii of $23.36-37.58$\,km than the other sources. It is thought that kHz QPOs reflect the motion of matter in orbit at some preferred radius $r$ in the accretion disk \citep{Miller98,Stella99a,Osherovich99,Lamb01,Zhang04}, where the magnetosphere-disk radius there $r_{\rm m}=r$ satisfies the equation (\ref{r_m_NS}): \\ $r_{\rm m}\approx7({\rm km})(\frac{B_{\rm s}}{10^8\,{\rm G}})^{4/7}(\frac{\dot{M}}{10^{18}\,{\rm g\,s^{-1}}})^{-2/7}$. \\ We suggest that Cir X-1 may have the large NS surface magnetic field $B_{\rm s}$, which will cause the large magnetosphere-disk radius, i.e. the large ones of the emission radii of kHz QPOs. It is not clear what physical parameters determine the appearance of the detectable kHz QPOs, nor why they only occur in the certain range of $1\sim\geq20$\,km away from the NS surface, and we suspect that it may be related to the accretion environment of LMXBs, which need the further study of the accretion state of the systems.
\item XTE J1807.4-294, 4U 0614+09, GX 5-1 and GX 340+0 show the larger range of the emission radii of kHz QPOs from $\sim$ 17\,km to $\sim$ 30\,km (see Table \ref{data_All} and Fig.\ref{r_Z}), implying these sources may have the more suitable physical environment to produce kHz QPOs. Less bright source XTE J1807.4-294 ($r=18.18\sim30.51$\,km) and 4U 0614+09 ($r=17.02\sim28.72$\,km) have the similar large range of the emission radii with the more bright source GX 5-1 ($r=16.90\sim29.69$\,km) and GX 340+0 ($r=17.97\sim27.18$\,km), which implies the emission condition of kHz QPOs may not sensitive to the mean accretion rate.
\item It can be seen from see Table \ref{data_All}, Table \ref{data_mass} and Fig.\ref{r_Atoll_Z_MSP} that the emission radius $r$ of kHz QPOs for all the sources are larger than the NS radius $R$ ($r>16$\,km and $R$ is assumed to be 15\,km, i.e. $r>R$), which means that the accretion matter that is related to the kHz QPOs is at the position at least one kilometer away from the NS surface. There may be some interpretations for these phenomena: One is the effect of the local strong magnetic spot \citep{Zhang06a} which expels the magnetosphere-disk not to approach the NS surface, so the accretion matter may at the position away from the NS surface. Another explanation is that the accretion plasma drops onto the stellar surface due to the instability when crossing the innermost stable circular orbit (ISCO) of NS, and the ISCO radius ($R_{\rm ISCO}=6GM/c^2$) is the inner boundary of the accretion matter, so the emission radius of the kHz QPOs should be larger than $R_{\rm ISCO}$.
    We try to test the ISCO effect scenario: Three sources in the sample, i.e. Cyg X-2, SAX J1808.4-3658 and XTE 1820-30, have both the detected twin kHz QPO frequencies and the measured NS masses (see Table \ref{data_mass}), which can be used to calculate both the NS radii and ISCO radii with the following equations:\\
    \\$\nu_1=629({\rm Hz})A^{-2/3}\nu_{2\rm k}^{5/3}\sqrt{1-\sqrt{1-(\frac{\nu_{2\rm k}}{1.85A})^{2/3}}}$,\\
    $A=\sqrt{\frac{M}{M_\odot}\frac{1}{R^3_6}}$,\\
    $R_{\rm ISCO}=6GM/c^2$.\\
    The results are shown in Table \ref{data_mass}, i.e. for Cyg X-2, $R\sim16.4$\,km and $R_{\rm ISCO}\sim15.12$\,km, for SAX J1808.4-3658, $R\sim18-20$\,km and $R_{\rm ISCO}\sim9-12$\,km, for XTE 1820-30, $R\sim14.5$\,km and $R_{\rm ISCO}\sim11.4$\,km. It can be seen that the ISCO radii of the three sources are all smaller than their inferred NS radii, so we suggest that the ISCO effect may not be the main reason to cause the emission radii of the kHz QPOs larger than the NS radii.
\item It can be seen from Table \ref{data_All}, Table \ref{data_mass} and Fig.\ref{r_Atoll_Z_MSP} that most Atoll and Z sources (19/23) are inferred to share the similar emission radii of kHz QPOs of $16-19$\,km. We try to probe the NS magnitude field by the kHz QPOs: The luminosities of Atoll sources are less than those of Z sources in 2-3 magnitude orders  in general, thus we assume that the accretion rates of Atoll sources are about 2 magnitude orders lower than those of Z sources in average. The magnetosphere-disk radius $r_{\rm m}$ where the kHz QPOs occur  satisfy the equation (\ref{r_m_NS}):\\
    $r_{\rm m}\approx7({\rm km})(\frac{B_{\rm s}}{10^8\,{\rm G}})^{4/7}(\frac{\dot{M}}{10^{18}\,{\rm g\,s^{-1}}})^{-2/7}$,\\
    the similar $r_{\rm m}(=16-19$\,km) and the different $\dot{M}$ will infer the NS surface magnetic field strengths of Atoll sources are about one magnitude order less than those of Z sources, which is consistent with the prediction of \citet{Zhang07c}. Source XTE J1701-462 has been observed both Atoll and Z behaviors \citep{Homan10}, which can go from one state to the other one in timescale of months, and \citet{Homan10} suggest this source has a NS surface magnetic field strength should not change in that timescale. We argue that XTE J1701-462 has a steady NS surface magnetic field and should not change in the short timescale, and ascribe the change of the Atoll and Z behaviors to the different magnetic field strengths at the different magnetosphere-disk radii: By assuming the NS magnetic field to be dipolar (see equation (\ref{NS_B})):\\ $B(r)=B_{\rm s}(\frac{R}{r})^3$,\\
    when the source has a instantaneously high accretion rate, the accretion disk will go closer to the NS surface and  shows a smaller $r_{\rm m}$, where the magnetic field there will be higher, then the source shows the Z source behavior. On the contract, the system will show the Atoll source behavior with the lower magnetic field at the magnetosphere-disk radius when it has a instantaneously low accretion rate. So we suggest that the Atoll and Z states of the source may reflect the magnetic field information at the different magnetosphere-disk radii.
\item Cir X-1 ($r_{\rm min}=23.36$\,km), Cyg X-2 ($r_{\rm min}=21.23$\,km) and SAX J1808.4-3658 ($r_{\rm min}=20$\,km) show the large values of the minimal emission radii of kHz QPOs compared with the other sources in the sample (see also Table \ref{data_All}, Table \ref{data_mass} and Fig.\ref{X_r}). As for Cir X-1 and Cyg X-2, we try to ascribe their larger ones of the minimal emission radii to share the stronger NS surface magnetic fields that make the averaged  magnetosphere-disk radii to be larger. For SAX J1808.4-3658, its inferred NS radius is slightly large ($R=18-20$\,km with the NS mass of $M=1-1.4\rm \,M_\odot$, see Table \ref{data_mass}), so its emission radii of kHz QPOs are relative large although it has the similar emission parameter $X$ values to the other sources ($r=R/X$).
\end{enumerate}

\section*{Acknowledgments}

We thank J. Wang, Z.B. Li and H.X. Yin for helpful discussions. This work is supported by the National Basic Research Program of China (2012CB821800 and 2009CB824800), the National Natural Science Foundation of China NSFC(11173034, 11173024, 11303047), the Science and Technology Foundation of Guizhou Province (Grant No.J[2015]2113), the Doctoral Starting up Foundation of Guizhou Normal University 2014 and the Innovation Team Foundation of the Education Department of Guizhou Province under Grant Nos. [2014]35.

\bsp

\label{lastpage}

\end{document}